\documentclass[11pt]{iopart}

\usepackage{bm}
\usepackage[english]{babel}
\usepackage{amssymb, ascmac}
\usepackage{cite}
\usepackage{color}

\newcommand{\be}{\beta}
\newcommand{\ka}{\kappa}
\newcommand{\bea}{\begin{eqnarray}}
\newcommand{\eea}{\end{eqnarray}}
\newcommand{\aeq}{&=&}
\newcommand{\aeqd}{&:= &}
\newcommand{\aeqe}{& =:&}
\newcommand{\aeqw}{& \stackrel{\mathrm{w}}{=} &}
\newcommand{\aeqap}{&\approx &}

\newcommand{\mL}{\mathcal{L}}

\newcommand{\tl}{\tilde}
\newcommand{\we}{\stackrel{\mathrm{w}}{=}}
\newcommand{\defe}{:=}
\newcommand{\ep}{\varepsilon}
\newcommand{\ga}{\gamma}
\newcommand{\Ga}{\Gamma}

\newcommand{\al}{\alpha}

\newcommand{\sig}{\sigma}
\newcommand{\f}{\frac}
\newcommand{\half}{\frac{1}{2}}
\newcommand{\pr}{\prime}
\newcommand{\dl}{\delta}

\newcommand{\lm}{\lambda}

\newcommand{\tot}{{\rm{tot}}}
\newcommand{\com}{\hspace{0.5mm}, \quad}
\newcommand{\la}{\label}

\newcommand{\no}{\nonumber}

\newcommand{\const}{\mbox{constant}}
\newcommand{\re}[1]{(\ref{#1})}
\newcommand{\res}[1]{\S \ref{#1}}
\newcommand{\hs}{\hspace}

\newcommand{\rd}[1]{\textcolor{black}{#1}}
\newcommand{\bl}[1]{\textcolor{black}{#1}}

\newcommand{\RM}[1]{{\rm{#1}}}
\newcommand{\p}{\partial}

\newcommand{\co}[1]{``{#1}''}
\newcommand{\bL}{\bm{\mathcal{L}}}

\begin{document}

\title[A Note on the Feynman Lectures on Gravitation]{A Note on the Feynman Lectures on Gravitation}

\author{Satoshi Nakajima}

\address{Graduate School of Pure and Applied Sciences, University of Tsukuba, 
1-1-1, Tennodai, Tsukuba 305-857, Japan}
\ead{subarusatosi@gmail.com}
\vspace{10pt}
\begin{indented}
\item[]\today
\end{indented}

\begin{abstract}

Following Feynman's lectures on gravitation, we consider the theory of the gravitational (massless spin-2) field in flat spacetime and present the third- and fourth-order Lagrangian densities for the gravitational field. 
In particular, we present detailed calculations for the third-order Lagrangian density. 
We point out that the expression for the third-order Lagrangian density which Feynman provided is 
\rd{not a solution of Feynman's condition that the third-order Lagrangian density must satisfy}. 
However, Feynman's third-order Lagrangian density gives the correct perihelion shift.

\end{abstract}

%
%
%
%
%

\section{Introduction}

\rd{General relativity can be viewed as the unique two-derivative nonlinear completion of a free massless spin‑2 field once locality, 
Lorentz invariance and a consistent coupling to a conserved stress tensor are imposed; see Wyss \cite{Wyss}, Deser \cite{1970}, 
and Wald \cite{Wald} for classic discussions of this \co{spin‑2 route} and its uniqueness (up to surface terms).
Early flat-spacetime field-theoretic approaches to gravity were developed by Gupta \cite{1954}. 
Kraichnan provided a special-relativistic derivation of generally covariant gravity \cite{1955} 
and analyzed the possibility of unequal gravitational and inertial masses in this framework \cite{1956}. 
Related formulations were also discussed by Thirring \cite{1961}, 
while Weinberg gave an S‑matrix argument leading to universal coupling and equality of gravitational and inertial mass 
for a massless spin‑2 particle \cite{1964}.
For pedagogical modern expositions of the flat-spacetime spin‑2 construction, see Ortín \cite{GS} and Janssen \cite{Janssen}.
Recent discussions have clarified subtleties of the iterative self-coupling (\co{bootstrap}) viewpoint 
and its assumptions; see, e.g., Deser's concise modern reformulation \cite{NEW‑DeserRedux} 
and the explicit bootstrapping analysis of Butcher, Hobson and Lasenby \cite{NEW‑ButcherHobsonLasenby}. 
A critical assessment of common bootstrap claims and related ambiguities is given by Padmanabhan \cite{NEW‑Padmanabhan}.
}

During his lectures on gravitational theory in 1962–1963, 
Feynman imagined Venusian scientists who knew field theory but not general relativity \cite{Feynman}. 
From the perspective of the Venusians, Feynman considered a theory of gravity in flat spacetime.
The gravitational field is represented as a symmetric tensor $h_{\mu\nu}$. 
Feynman first considered the quadratic Lagrangian density term in $h_{\mu\nu}$ and derived the Fierz-Pauli Lagrangian density \cite{1939}.
Next, Feynman derived the equation of motion for a point mass in the gravitational field and 
used it to derive the equation for the divergence of the energy-momentum tensor for the point mass system.
Based on this, Feynman derived the condition that the third-order Lagrangian density term in $h_{\mu\nu}$ must satisfy. 
\rd{This condition is the perturbative form of the (nonlinear) Bianchi identity. }
However, the expression for the third-order Lagrangian density that Feynman provided, \rd{$\bL^{(3)}_\RM{Feynman}$, does not satisfy
the condition and 
\bea
4\ka (\bL^{(3)}_\RM{Feynman}- \bL_\RM{E}^{(3)}) 
\aeqw -h_{\al\be}\p_\ga h^{\al\dl}\p_\dl h^{\be\ga}+h_{\al\be}\p^\be h^{\al\ga}\p^\dl h_{\dl\ga} \stackrel{\mathrm{w}}{\ne} 0
\eea
holds. 
Here, $\ka$ is the Einstein constant and $\bL_\RM{E}^{(3)}$ is the Einstein's third-order Lagrangian density, 
which satisfies Feynman's condition. 
$A \we B$ means that there exists $C^\mu$ such that $A=B+\p_\mu C^\mu$.
}

\rd{In this note, we assume the following axioms:
\begin{enumerate}
  \item Locality and Lorentz invariance.
  \item At most two derivatives in field equations.
  \item The principle of equivalence (universal coupling to the conserved stress tensor). 
  \item The linear Bianchi identity for the second-order Lagrangian density. 
  \item The Bianchi identity. 
\end{enumerate}
}

The structure of this note is as follows. 
First, we consider a \bl{point-mass system} coupled to the gravitational field (\res{s_2}). 
Next, we study the action of the gravitational field (\res{s_4}). 
In \res{A_3}, we present detailed calculations for the third-order Lagrangian density. 
In \res{A_4}, we study the fourth-order Lagrangian density. 
In \res{s_6}, we explain the perihelion shift based on the Feynman lectures \cite{Feynman}. 
In Appendix \ref{A_B}, we calculate third-order Lagrangian densities. 

\rd{This note is intended to: 
\begin{itemize}
  \item  Specific corrections to the widely read Feynman Lectures
(Educational value).
 \item Visualizing the modern understanding of GR's uniqueness from spin-2 using Feynman's example
(Conceptual value).
\item Organization of explicit third- and fourth-order Lagrangians (in a reference-friendly form)
(Technical reference value).
\end{itemize}
}

\section{Point mass system} \la{s_2}

We consider the Minkowski spacetime. The metric is $\eta_{\mu\nu} = \RM{diag}(-1,1,1,1)$.
The gravitational field is represented as a symmetric tensor $h_{\mu\nu}$.

We suppose that the action for a \bl{point-mass system} and the gravitational field is given by
\bea
S \aeq S_\RM{particle} + S_\RM{int}+S_\RM{Gravity} ,\\
S_\RM{particle} \aeq \sum_a \f{m_a}{2} \int d\lm_a \ \Big[e_a(\lm_a)\eta_{\mu\nu}\f{dz^\mu_a}{d\lm_a}\f{dz^\nu_a}{d\lm_a} 
-\f{c^2}{e(\lm_a)} \Big],\\
S_\RM{int} \aeq \sum_a \f{g_a}{2} \int d\lm_a \ e_a(\lm_a)h_{\mu\nu}(z_a)\f{dz^\mu_a}{d\lm_a}\f{dz^\nu_a}{d\lm_a}.
\eea
Here, $m_a$ is a mass of particle $a$ and 
$g_a$ is a coupling constant. 
\bl{$z_a^\mu$ is the space-time coordinate of particle $a$.} 
$\lm_a$ is a parameter and $e_a$ is an auxiliary field. 
$S_\RM{Gravity}$ is the action of the gravitational field. 
$S_\RM{particle}$ and $S_\RM{int}$ are invariant under a transformation $\lm_a \to \lm_a^\pr$ and $e_a \to e_a^\pr=\f{d\lm_a^\pr}{d\lm_a}e_a$. 
We denote by $\tau_a$ the parameter for which $e_a$ becomes 1.
Then, we have
\bea
S_\RM{particle} \aeq \sum_a \f{m_a}{2} \int d\tau_a \ \Big[\eta_{\mu\nu}\f{dz^\mu_a}{d\tau_a}\f{dz^\nu_a}{d\tau_a} 
-c^2 \Big],\\
S_\RM{int} \aeq \sum_a \f{g_a}{2} \int d\tau_a \ h_{\mu\nu}(z_a)\f{dz^\mu_a}{d\tau_a}\f{dz^\nu_a}{d\tau_a}.
\eea
We denote the first term of $\tl S_\RM{particle}$ by $S_\RM{particle}$.
The second term of $S_\RM{particle}$ does not contribute to the variation. 
The action of the particles can be rewritten as
\bea
S_p \aeqd \tl S_\RM{particle} + S_\RM{int}
= \sum_a \f{m_a}{2} \int d\tau_a \ \Big(\eta_{\mu\nu}+\f{g_a}{m_a}h_{\mu\nu}(z_a) \Big)\f{dz^\mu_a}{d\tau_a}\f{dz^\nu_a}{d\tau_a} \no\\
\aeq \sum_a \f{m_a}{2} \int d\tau_a \ g_{\mu\nu}^{(a)}(z_a)\f{dz^\mu_a}{d\tau_a}\f{dz^\nu_a}{d\tau_a},
\eea
where
\bea
g_{\mu\nu}^{(a)} \aeqd \eta_{\mu\nu}+\f{g_a}{m_a}h_{\mu\nu} .
\eea
The variation is given by
\bea
\dl S_p 
\aeq \sum_a \f{m_a}{2}\int d\tau_a \  \dl z_a^\lm \cdot (-2)\Big( \half \big[-\p_\lm g_{\mu\nu}^{(a)}+\p_\mu g_{\lm\nu}^{(a)}+ \p_\nu g_{\lm\mu}^{(a)} \big]
\f{dz^\mu_a}{d\tau_a}\f{dz^\nu_a}{d\tau_a} \no\\
&&+g_{\lm\nu}^{(a)}(z_a)\f{d^2z^\nu_a}{d\tau_a^2} \Big) .
\eea
Then, the equation of motion of particle $a$ is given by
\bea
\hs{-5mm}\Big(m_a\eta_{\lm\nu}+g_a h_{\lm\nu}(z_a) \Big)\f{d^2z^\nu_a}{d\tau_a^2}
+\half g_a \big[-\p_\lm h_{\mu\nu}+\p_\mu h_{\lm\nu}+ \p_\nu h_{\lm\mu} \big]\f{dz^\mu_a}{d\tau_a}\f{dz^\nu_a}{d\tau_a} \aeq 0 . \la{eqm_a}
\eea
According to the principle of equivalence, the ratio $g_a/m_a$ does not depend on the type of particle.
Then, we set $g_a=m_a$. 
\re{eqm_a} becomes
\bea
g_{\lm\nu}(z_a)\f{d^2z^\nu_a}{d\tau_a^2}+\Ga_{\lm\mu\nu}(z_a)\f{dz^\mu_a}{d\tau_a}\f{dz^\nu_a}{d\tau_a} \aeq 0 ,\la{p_eom}
\eea
where
\bea
g_{\mu\nu} \aeqd \eta_{\mu\nu}+h_{\mu\nu} ,\\
\Ga_{\lm\mu\nu} \aeqd \half \big[-\p_\lm h_{\mu\nu}+\p_\mu h_{\lm\nu}+ \p_\nu h_{\lm\mu} \big].
\eea
The Euler-Lagrange equation of $e_a$ for $e_a=1$ is given by
\bea
g_{\mu\nu}(z_a)\f{dz^\mu_a}{d\tau_a}\f{dz^\nu_a}{d\tau_a} \aeq -c^2. \la{e_a_eom}
\eea

We define the energy-momentum tensor of the particles as
\bea
\bm{T}^{\mu\nu}_\RM{(p)}(x) \aeqd \sum_a  m_a \int d\tau_a \ \dl^4(x-z_a)\f{dz^\mu_a}{d\tau_a}\f{dz^\nu_a}{d\tau_a}.
\eea
Then, $S_\RM{int} $ can be rewritten as
\bea
S_\RM{int} \aeq \int d^4x \ \half h_{\mu\nu}(x)\bm{T}^{\mu\nu}_\RM{(p)}(x).
\eea
Using
\bea
\p_\nu \bm{T}^{\mu\nu}_\RM{(p)} \aeq \sum_a  m_a \int d\tau_a \ \p_\nu \dl^4(x-z_a)\f{dz^\mu_a}{d\tau_a}\f{dz^\nu_a}{d\tau_a} \no\\
\aeq \sum_a  m_a \int d\tau_a \ (-1)\f{d\dl^4(x-z_a)}{d\tau_a}\f{dz^\mu_a}{d\tau_a} \no\\
\aeq \sum_a  m_a \int d\tau_a \ \dl^4(x-z_a) \f{d^2z^\mu_a}{d\tau_a^2}
\eea
and \re{p_eom}, we have
\bea
g_{\lm\mu}\p_\nu \bm{T}^{\mu\nu}_\RM{(p)} \aeq \sum_a  m_a \int d\tau_a \ \dl^4(x-z_a) g_{\lm\mu}(z_a)\f{d^2z^\mu_a}{d\tau_a^2} \no\\
\aeq \sum_a  m_a \int d\tau_a \ \dl^4(x-z_a) \Big[-\Ga_{\lm\mu\nu}(z_a)\f{dz^\mu_a}{d\tau_a}\f{dz^\nu_a}{d\tau_a} \Big] \no\\
\aeq -\Ga_{\lm\mu\nu}(x)\sum_a  m_a \int d\tau_a \ \dl^4(x-z_a) \f{dz^\mu_a}{d\tau_a}\f{dz^\nu_a}{d\tau_a} \no\\
\aeq -\Ga_{\lm\mu\nu}(x)\bm{T}^{\mu\nu}_\RM{(p)}(x). \la{eq_T_p}
\eea
We denote matter fields system as $S_\RM{matter}$ and define $\bm{T}^{\mu\nu}_\RM{(m)} $ as
\bea
\dl S_\RM{matter} \aeq \int d^4 x \ \dl h_{\mu\nu}(x) \half \bm{T}^{\mu\nu}_\RM{(m)} .
\eea
We suppose that the total energy-momentum tensor $\bm{T}^{\mu\nu} \defe \bm{T}^{\mu\nu}_\RM{(p)}+\bm{T}^{\mu\nu}_\RM{(m)}$ 
also satisfies 
\bea
g_{\lm\mu}\p_\nu \bm{T}^{\mu\nu} \aeq -\Ga_{\lm\mu\nu}\bm{T}^{\mu\nu} .\la{eq_T}
\eea

\section{Action of the gravitational field: Venusian calculations} \la{s_4}

In \res{s_4_general}, we consider the action of gravity. 
First, we study the second-order Lagrangian density term in $h_{\mu\nu}$ (\res{s_4_2}).
Next, we study the third-order Lagrangian density $\bL^{(3)}$ (\res{A_3}). 
In \res{s_4_4}, we point out that the third-order Lagrangian density provided by Feynman is 
\rd{not a solution of Feynman's condition that the third-order Lagrangian density must satisfy}.

In the following, we set $c=1$.

\subsection{General theory} \la{s_4_general}

We expand the action of the gravitational field $S_\RM{Gravity} $ as
\bea
S_\RM{Gravity} \aeq \sum_{n=2}^\infty S^{(n)} \com S^{(n)} =\int d^4x \ \bL^{(n)}.
\eea
Here, $\bL^{(n)}$ is $n$-th-order term in $h_{\mu\nu}$. 
We introduce $\bm{\chi}^{\mu\nu}$ and $\bm{\chi}_{(n)}^{\mu\nu}$ as
\bea
\dl S_\RM{Gravity} \aeq -\half \int d^4x \ \dl h_{\mu\nu}\bm{\chi}^{\mu\nu} ,\\
\dl S^{(n)} \aeq -\half \int d^4x \ \dl h_{\mu\nu}\bm{\chi}_{(n-1)}^{\mu\nu}.
\eea
Then, $\bm{\chi}^{\mu\nu} =\sum_{n=1}^\infty\bm{\chi}_{(n)}^{\mu\nu} $ holds. 
The Euler-Lagrange equation of gravity is given by
\bea
\bm{\chi}^{\mu\nu} = \bm{T}^{\mu\nu} .\la{eq_chi}
\eea
We assume that $\bm{\chi}_{(1)}^{\mu\nu} $ and $\bm{\chi}^{\mu\nu}$ satisfy
\bea
\p_\nu \bm{\chi}_{(1)}^{\mu\nu} =0 ,\la{cond_0}\\ 
g_{\lm\mu}\p_\nu \bm{\chi}^{\mu\nu}\aeq -\Ga_{\lm\mu\nu}\bm{\chi}^{\mu\nu} \la{eq_T_chi}
\eea
without using \re{eq_chi}. 
\rd{
\re{cond_0} corresponds to the gauge invariance of $\bm{\chi}_{(1)}^{\mu\nu} $ under 
$h_{\al\be} \to h_{\al\be}+\p_\al \chi_\be + \p_\be \chi_\al$. 
Here, $\chi_\be$ is an infinitesimal vector. 
\re{cond_0} is the linear Bianchi identity. 
}
\re{eq_T_chi} has the same form as \re{eq_T} \rd{and corresponds to the Bianchi identity}. 
The above two equations lead to 
\bea
(\eta_{\lm\mu}+h_{\lm\mu})\sum_{n=2}^\infty \p_\nu \bm{\chi}_{(n)}^{\mu\nu} +\Ga_{\lm\mu\nu}\sum_{n=1}^\infty \bm{\chi}_{(n)}^{\mu\nu}  \aeq 0
\eea
and 
\bea
\eta_{\lm\mu}\p_\nu \bm{\chi}_{(2)}^{\mu\nu} \aeq -\Ga_{\lm\mu\nu}\bm{\chi}_{(1)}^{\mu\nu} , \la{cond_2}\\
\eta_{\lm\mu}\p_\nu \bm{\chi}_{(n+1)}^{\mu\nu} \aeq -\Ga_{\lm\mu\nu}\bm{\chi}_{(n)}^{\mu\nu}
-h_{\lm\mu}\p_\nu \bm{\chi}_{(n)}^{\mu\nu} \ \ (n=2,3,\cdots) .\la{cond_3}
\eea

The candidate of $\bL^{(2)}$ is given by
\bea
\bL^{(2)} \aeq \half \Big[a_1 \p_\al h_{\mu\nu} \p^\al h^{\mu\nu}
+a_2\p_\al h_\mu^{\  \nu}\p_\nu h^{\mu\al}
+a_3(\p h)^{\mu}\p_\mu h \no\\
&&+a_4\p^\mu h \p_\mu h +a_5(\p h)^{\mu}(\p h)_\mu \Big], \la{intro_a_i}
\eea
where $h\defe h^\mu_{\ \mu}$ and $(\p h)^{\nu} \defe \p_\mu h^{\mu\nu} $. 
Because $(\p h)^{\mu}(\p h)_\mu \we \p_\al h_\mu^{\  \nu}\p_\nu h^{\mu\al}$, 
we can set $a_5=0$. 
Here, $A \we B$ means that there exists $C^\mu$ such that $A=B+\p_\mu C^\mu$. 
From \bl{\re{cond_0}}, the ratios $a_2/a_1$, $a_3/a_1$, and $a_4/a_1$ are determined.
$a_1$ is determined from \re{eq_chi} in the Newtonian limit. 
$\bL^{(3)}$ is determined from \re{cond_2}, which is equivalent to (3.197) in Ref.\cite{GS} and (4.20) in Ref.\cite{Wyss}.
The candidate of $\bL^{(3)}$ has 16 terms. 
We determine $\bL^{(3)}$ in \res{A_3}. 
$\bL^{(4)}$, $\bL^{(5)},\cdots $ are determined by \re{cond_3}. 
The candidates of $\bL^{(4)}$, $\bL^{(5)}$, $\bL^{(6)}$, $\bL^{(7)}$, and $\bL^{(8)}$ have
43, 93, 187, 344, and 607 terms, respectively. 
We determine $\bL^{(4)}$ in \res{A_4}. 

The Einstein-Hilbert Lagrangian density is equivalent to the Einstein Lagrangian density $\bm{\mL}_\RM{E}$ 
defined by 
\bea
\bm{\mL}_\RM{E} \aeqd \f{1}{2\ka}\sqrt{-\det(g_{\mu\nu})}G \com 
G \defe g^{\mu\nu}\Big[  \Ga^\rho_{\ \ga\nu}\Ga^\ga_{\ \mu\rho}-\Ga^\rho_{\ \ga\rho}\Ga^\ga_{\ \mu\nu} \Big]. 
\eea
Here, $\ka$ is the Einstein constant and $\Ga^\rho_{\ \mu\nu}\defe g^{\rho\lm}\Ga_{\lm\mu\nu}$ where
$g^{\mu\nu}$ is the inverse matrix of $g_{\mu\nu}$. 
Then, 
\bea
\bm{\chi}_\RM{E}^{\mu\nu} \aeqd -2\Big(\f{\p \bm{\mL}_\RM{E}}{\p h_{\mu\nu}}-\p_\sig \f{\p \bm{\mL}_\RM{E}}{\p (\p_\sig h_{\mu\nu})}\Big)
\eea
satisfies \re{eq_T_chi} identically. 
If we expand $\bm{\mL}_\RM{E}$ as $\bm{\mL}_\RM{E}=\bm{\mL}_\RM{E}^{(2)}+\bm{\mL}_\RM{E}^{(3)}+\cdots$, 
\bea
\bL^{(n)} \aeqw \bL_\RM{E}^{(n)}
\eea
should be satisfied.

\subsection{Second-order Lagrangian density} \la{s_4_2}

We determine $\{a_i\}_{i=1}^4$ of \re{intro_a_i}. 
First, we have
\bea
\bm{\chi}_{(1)}^{\mu\nu} \aeq 2a_1\Box h^{\mu\nu}+a_2(\p^\mu (\p h)^{\nu}+\p^\nu (\p h)^{\mu}) \no\\
&&+a_3[\p^\mu \p^\nu h +\eta^{\mu\nu}(\p \p h)  ]+2a_4 \eta^{\mu\nu}\Box h,
\eea
where $(\p \p h) \defe \p_\al \p_\be h^{\al\be}$ and $\Box\defe \p^\mu \p_\mu$. 
The above equation leads to
\bea
\p_\nu \bm{\chi}_{(1)}^{\mu\nu} \aeq 2a_1\Box (\p h)^{\mu}+a_2(\p^\mu (\p \p h)+\Box (\p h)^{\mu}) \no\\
&&+a_3(\p^\mu \Box h+\p^\mu (\p \p h))+2a_4 \p^\mu \Box h .
\eea
Because of \bl{\re{cond_0}}, we have
\bea
2a_1+a_2 = 0 \com a_2+a_3 = 0 \com \bl{a_3+2a_4} = 0,
\eea
namely, $a_2 = -2a_1$, $a_3=2a_1$, and $a_4 = -a_1$.
Then, we have
\bea
\bm{\chi}_{(1)}^{\mu\nu} \aeq 2a_1 \Big[ \Box h^{\mu\nu}-(\p^\mu (\p h)^{\nu}+\p^\nu (\p h)^{\mu}) \no\\
&&+[\p^\mu \p^\nu h +\eta^{\mu\nu}(\p \p h) ]- \eta^{\mu\nu}\Box h \Big] , \la{chi_1}\\
\bL^{(2)} \aeq a_1 \Big[ \half \p_\al h_{\mu\nu} \p^\al h^{\mu\nu}-\p_\al h_\mu^{\  \nu}\p_\nu h^{\mu\al}
+(\p h)^{\mu}\p_\mu h -\half \p^\mu h \p_\mu h \Big].
\eea
In the Newtonian limit, \re{eq_chi} leads to $a_1 = -\f{1}{4\ka}$. 
Then, $\bL^{(2)}=\bL^{(2)}_\RM{E}$ holds. 
$\bL^{(2)}$ is the Fierz-Pauli Lagrangian density \cite{1939}.

\subsection{Third-order Lagrangian density} \la{A_3}

We determine $\bL^{(3)}$. 
The candidate of  $\bL^{(3)}$ is given by 
\bea
\bL^{(3)} \aeq \sum_{\sig \in S_4} g_\sig (\sig(1)\sig(2)\sig(3)\sig(4)) ,
\eea
where 
\bea
(i_1i_2i_3i_4) \aeqd h^{\mu_{i_1}}_{\ \ \mu_1}\p_{\mu_2}h^{\mu_{i_2}}_{\ \ \mu_3}\p^{\mu_{i_3}}h^{\mu_{i_4}}_{\ \ \mu_4} 
\eea
and $S_4$ is the fourth-order permutation group. 
Because of
\bea
(1342)\aeq (1234),\ (3214)=(2341),\ (3412)=(2431),\ (4213)=(2143) , \no\\
(4123)\aeq (3421),\ (4231)=(3142),\ (4312)=(2134),\ (4321)=(3124) ,
\eea
there are 16 independent terms. 
$\bL^{(3)}$ is given by
\bea
\bL^{(3)} \aeq g_1 h\p_\al h \p^\al h+g_2 h\p_\ga h^{\al\be}\p^\ga h_{\al\be} +g_3h\p_\ga h^{\al\be}\p_\be h^{\ga}_{\ \al}
+g_4h_{\al\be}\p^\al h \p^\be h \no\\
&&+ g_5 h_{\al\be}\p^\al h^\dl_{\ \ga}\p^\be h^\ga_{\ \dl}+g_6h_{\al\be}\p^\al h^{\ga\dl}\p_\ga h^\be_{\ \dl} 
+g_7 h_{\al\be}\p_\ga h^{\al\dl}\p^\ga h^\be_{\ \dl} \no\\
&&+g_8 h_{\al\be}\p_\ga h^{\al\dl}\p_\dl h^{\be\ga}
+g_9 h_{\al\be}\p_\ga h \p^\al h^{\be\ga} 
+g_{10} h_{\al\be}\p_\ga h \p^\ga h^{\al\be} \no\\
&&+g_{11}h(\p h)^{\al}\p_\al h 
+ g_{12}h_{\al\be}\p^\be h^{\al\ga}(\p h)_{\ga} 
+g_{13}h_{\al\be}\p^\al h (\p h)^{\be} \no\\
&&+g_{14} h_{\al\be}\p_\ga h^{\al\be}(\p h)^{\ga}
+g_{15}h(\p h)_{\al}(\p h)^{\al} + g_{16}h_{\al\be}(\p h)^{\al}(\p h)^{\be} \no\\
\aeqe \sum_{i=1}^{16}g_i[i] .\la{L_3_General}
\eea
In the following, we calculate
\bea
\bm{\chi}^{\mu\nu}_{(2)} \aeq -2\Big( \f{\p \bL^{(3)} }{\p h_{\mu\nu}}-\p_\lm \f{\p \bL^{(3)} }{\p (\p_\lm h_{\mu\nu})} \Big) 
=: \sum_{i=1}^{16}g_i \bm{\chi}^{\mu\nu}_{[i]}
\eea
and $(\p \bm{\chi}_{[i]})_\lm \defe \eta_{\lm\mu}\p_\nu \bm{\chi}^{\mu\nu}_{[i]}$. 
\re{cond_2} can be rewritten as
\bea
\sum_{i=1}^{16}g_i (\p \bm{\chi}_{[i]})_\mu \aeq -\Ga_{\mu\al\be}\bm{\chi}_{(1)}^{\al\be} =:V_\mu .\la{cond_2_re}
\eea
Using \re{chi_1}, we have 
\bea
\hs{-5mm}V_\mu/g \aeq -2\p_\mu h_{\al\be}\Box h^{\al\be}+4\p_\mu h_{\al\be}\p^\al (\p h)^\be -2\p_\mu h_{\al\be} \p^\be\p^\al h
+2\p_\mu h \Box h  \no\\
&&-2 \p_\mu h (\p \p h) +4\p_\al h_{\mu\be}\Box h^{\al\be} 
-4\p_\al h_{\mu\be}\p^\al (\p h)^\be \no\\
&&-4\p_\al h_{\mu\be}\p^\be (\p h)^\al 
+4\p_\al h_{\mu\be} \p^\be\p^\al h 
-4(\p h)_\mu \Box h+4(\p h)_\mu(\p \p h).
\eea
Here, $g\defe 1/(8\ka)$. 

$\{ [i] \}_{i=1}^{16}$ are not independent. 
We consider a Lorentz scalar quantity $a\p_\mu b \p_\nu c$. 
The superscripts $\mu$ and $\nu$ are also included in $a$, $b$, and $c$. 
Using
\bea
a\p_\mu b \p_\nu c \aeqw -\p_\nu(a\p_\mu b)c \no\\
\aeq -\p_\nu a\p_\mu bc- a\p_\nu \p_\mu bc \no\\
\aeqw -\p_\nu a\p_\mu bc+ \p_\mu (ac)\p_\nu b\no\\
\aeq -c\p_\nu a\p_\mu b + c\p_\mu a \p_\nu b +a \p_\mu c \p_\nu b , \la{weak_relation}
\eea
we have
\bea
[3] \aeqw -[9]+[13]+[15] \com  [6] \we  -[8]+[16]+[12]. \la{weak}
\eea
Applying \re{weak_relation} to $[11]$ and $[14]$ yields only trivial expressions ($[11] \we [11]$ and $[14] \we [14]$).
We \bl{do} not need to consider $h\p_\mu b \p^\mu c$, $h^{\al\be}\p_\mu b \p^\mu c$, and $h^{\mu\nu}\p_\mu b \p_\nu c$ type terms
because of 
\bea
a^{\mu\nu}\p_\mu b \p_\nu c \aeqw -c\p_\nu a^{\mu\nu}\p_\mu b + c\p_\mu a^{\mu\nu} \p_\nu b +a^{\mu\nu} \p_\mu c \p_\nu b \no\\
\aeq a^{\mu\nu} \p_\nu c \p_\mu b
\eea
for $a^{\mu\nu}=a^{\nu\mu}$. 

We have 
\bea
\bm{\chi}^{\mu\nu}_{[1]} 
\aeq \eta^{\mu\nu}[2\p_\al h \p^\al h + 4h \Box h] ,\\
\bm{\chi}^{\mu\nu}_{[2]} 
\aeq -2\eta^{\mu\nu}\p_\ga h^{\al\be}\p^\ga h_{\al\be}+4\p_\ga h\p^\ga h^{\mu\nu}+4h\Box h^{\mu\nu} ,\\
\bm{\chi}^{\mu\nu}_{[3]}
\aeq -2\eta^{\mu\nu}\p_\ga h^{\al\be}\p_\be h^{\ga}_{\ \al}+2\p_\ga h\p^\nu h^{\ga\mu}+2\p_\ga h\p^\mu h^{\ga\nu} \no\\
&&+2h\p^\mu (\p h)^{\nu}+2h\p^\nu (\p h)^{\mu} ,\\
\bm{\chi}^{\mu\nu}_{[4]} 
\aeq -2\p^\mu h \p^\nu h+4\eta^{\mu\nu}[(\p h)^{\al}\p_\al h +h^{\al\be}\p_\al \p_\be h] ,\\
\bm{\chi}^{\mu\nu}_{[5]} 
\aeq -2\p^\mu h^{\al\be}\p^\nu h_{\al\be}+4(\p h)^{\al}\p_\al h^{\mu\nu}+4 h^{\al\be}\p_\al \p_\be h^{\mu\nu} ,\\
\bm{\chi}^{\mu\nu}_{[6]} 
\aeq -\p^\mu h^{\ga\dl}\p_\ga h^\nu_{\ \dl}- \p^\nu h^{\ga\dl}\p_\ga h^\mu_{\ \dl}
+(\p h)^{\al}\p^\mu h_{\al}^{\ \nu}+(\p h)^{\al}\p^\nu h_{\al}^{\ \mu} \no\\
&&+ h^{\al\be}\p_\al \p^\mu h_{\be}^{\ \nu} 
+ h^{\al\be}\p_\al \p^\nu h_{\be}^{\ \mu} 
+2\p_\ga h^{\mu\al} \p_\al h^{\ga\nu} \no\\
&&+ h^{\mu\al} \p_\al (\p h)^{\nu} +h^{\nu\al} \p_\al (\p h)^{\mu} ,\\
\bm{\chi}^{\mu\nu}_{[7]} 
\aeq \bl{2\p_\ga h^{\mu}_{\ \be} \p^\ga h^{\be\nu}}+2h^{\mu}_{\ \be} \Box h^{\be\nu}+2h^{\nu}_{\ \be} \Box h^{\be\mu} ,\\
\bm{\chi}^{\mu\nu}_{[8]}
\aeq -2\p_\ga h^{\mu\dl}\p_\dl h^{\nu\ga}+2\p_\ga h^{\mu}_{\ \be}\p^\nu h^{\be\ga}+2\p_\ga h^{\nu}_{\ \be}\p^\mu h^{\be\ga} \no\\
&&+2 h^{\mu}_{\ \be}\p^\nu (\p h)^{\be}+2 h^{\nu}_{\ \be}\p^\mu (\p h)^{\be} ,\\
\bm{\chi}^{\mu\nu}_{[9]} 
\aeq -\p_\ga h \p^\mu h^{\nu\ga}-\p_\ga h \p^\nu h^{\mu\ga}+2\eta^{\mu\nu}[\p_\ga h_{\al\be} \p^\al h^{\be\ga}+h^{\al\be} \p_\al (\p h)_\be ] \no\\
&&+(\p h)^{\mu}\p^\nu h+(\p h)^{\nu}\p^\mu h+ h^{\al\mu}\p_\al \p^\nu h +h^{\al\nu}\p_\al \p^\mu h ,\\
\bm{\chi}^{\mu\nu}_{[10]}
\aeq 2\eta^{\mu\nu}[\p_\ga h_{\al\be}\p^\ga h^{\al\be}+h_{\al\be}\Box h^{\al\be}]+2h^{\mu\nu}\Box h , \\
\bm{\chi}^{\mu\nu}_{[11]} 
\aeq 2\p^\mu h\p^\nu h +2h\p^\mu \p^\nu h+2\eta^{\mu\nu}h(\p \p h) ,\\
\bm{\chi}^{\mu\nu}_{[12]} 
\aeq -\p^\mu h^{\nu\ga}(\p h)_{\ga}-\p^\nu h^{\mu\ga}(\p h)_{\ga}+2(\p h)^\mu (\p h)^\nu +h^{\mu\be}\p_\be (\p h)^{\nu} \no\\
&&+ h^{\nu\be}\p_\be (\p h)^{\mu} 
+\p^\mu h^{\al\be}\p_\be h^\nu_{\ \al}+\p^\nu h^{\al\be}\p_\be h^\mu_{\ \al} \no\\
&&+h^{\al\be}\p_\be \p^\mu h^\nu_{\ \al}
+h^{\al\be}\p_\be \p^\nu h^\mu_{\ \al} ,\\
\bm{\chi}^{\mu\nu}_{[13]} 
\aeq -\p^\mu h (\p h)^{\nu}-\p^\nu h (\p h)^{\mu}+2\eta^{\mu\nu} [(\p h)^{\al} (\p h)_{\al}+h^{\al\be} \p_\al (\p h)_{\be}] \no\\
&&+\p^\mu h^{\al\nu}\p_\al h+\p^\nu h^{\al\mu}\p_\al h 
+h^{\al\nu}\p^\mu \p_\al h+ h^{\al\mu}\p^\nu \p_\al h,\\
\bm{\chi}^{\mu\nu}_{[14]} 
\aeq 2h^{\mu\nu}(\p \p h)+2\p^\mu h_{\al\be}\p^\nu h^{\al\be}  +2h_{\al\be}\p^\mu \p^\nu h^{\al\be} ,\\
\bm{\chi}^{\mu\nu}_{[15]} 
\aeq -2\eta^{\mu\nu}(\p h)_{\al}(\p h)^{\al}+2\p^\mu h(\p h)^{\nu}+2\p^\nu h(\p h)^{\mu} \no\\
&&+2 h\p^\mu(\p h)^{\nu}+2 h\p^\nu(\p h)^{\mu} ,\\
\bm{\chi}^{\mu\nu}_{[16]} 
\aeq  -2 (\p h)^{\mu}(\p h)^{\nu}+2\p^\mu h^{\nu\al}(\p h)_\al +2\p^\nu h^{\mu\al}(\p h)_\al \no\\
&&+2 h^{\nu\al}\p^\mu(\p h)_\al +2 h^{\mu\al}\p^\nu(\p h)_\al
\eea
and 
\bea
(\p \bm{\chi}_{[1]})_\mu 
\aeq 4\p_\al h \p_\mu \p^\al h
+ 4\p_\mu h \Box h+4h \p_\mu \Box h ,\\
(\p \bm{\chi}_{[2]})_\mu 
\aeq -4\p_\ga h_{\al\be}\p_\mu \p^\ga h^{\al\be}
+4\p_\nu \p_\ga h\p^\ga h_\mu^{\ \nu} \no\\
&&+4\p_\ga h\p^\ga (\p h)_\mu 
+4\p_\nu h\Box h_\mu^{\ \nu} +4h\Box (\p h)_\mu ,\\
(\p \bm{\chi}_{[3]})_\mu 
\aeq -2\p_\mu \p_\ga h^{\al\be}\p_\be h^{\ga}_{\ \al}
-2 \p_\ga h^{\al\be}\p_\mu \p_\be h^{\ga}_{\ \al}
+2\p_\nu \p_\ga h\p^\nu h^{\ga}_{\ \mu} \no\\
&&+2 \p_\ga h\Box h^{\ga}_{\ \mu} 
+2\p_\nu \p_\ga h\p_\mu h^{\ga\nu} 
+4 \p_\ga h\p_\mu (\p h)^{\ga} \no\\
&&+2h\p_\mu (\p \p h) 
+2\p_\nu h\p^\nu (\p h)_{\mu}+2 h \Box (\p h)_{\mu} , \\
(\p \bm{\chi}_{[4]})_\mu 
\aeq -2\p_\nu \p_\mu h \p^\nu h -2\p_\mu h \Box h +4\p_\mu (\p h)^{\al}\p_\al h \no\\
&&+4(\p h)^{\al}\p_\mu \p_\al h 
+4\p_\mu h^{\al\be}\p_\al \p_\be h+4h^{\al\be}\p_\mu \p_\al \p_\be h ,\\
(\p \bm{\chi}_{[5]})_\mu 
\aeq -2\p_\nu \p_\mu h^{\al\be}\p^\nu h_{\al\be}-2\p_\mu h^{\al\be}\Box h_{\al\be}+4\p_\nu (\p h)^{\al}\p_\al h_\mu^{\ \nu} \no\\
&&+4(\p h)^{\al}\p_\al (\p h)_\mu 
+4 \p_\nu h^{\al\be}\p_\al \p_\be h_\mu^{\ \nu}+4 h^{\al\be}\p_\al \p_\be (\p h)_\mu , \\
(\p \bm{\chi}_{[6]})_\mu 
\aeq  -\p_\nu\p_\mu h^{\ga\dl}\p_\ga h^\nu_{\ \dl}
- \Box h^{\ga\dl}\p_\ga h_{\mu\dl} - \p^\nu h^{\ga\dl}\p_\nu \p_\ga h_{\mu\dl} \no\\
&&+ (\p h)^{\al}\p_\mu (\p h)_{\al}
+\p_\nu (\p h)^{\al}\p^\nu h_{\al\mu} 
+ (\p h)^{\al}\Box h_{\al\mu} \no\\
&&+ \p_\nu h^{\al\be}\p_\al \p_\mu h_{\be}^{\ \nu} + h^{\al\be}\p_\al \p_\mu (\p h)_{\be} 
+ \p_\nu h^{\al\be}\p_\al \p^\nu h_{\be\mu} \no\\
&&+h^{\al\be}\p_\al \Box h_{\be\mu} 
+2\p_\nu \p_\ga h_{\mu}^{\ \al} \p_\al h^{\ga\nu} +3 \p_\ga h_{\mu}^{\ \al} \p_\al (\p h)^{\ga} \no\\
&&+h_{\mu}^{\ \al} \p_\al (\p \p h) 
+h^{\nu\al} \p_\nu \p_\al (\p h)_{\mu}+(\p h)^\al \p_\al (\p h)_\mu ,\\
(\p \bm{\chi}_{[7]})_\mu 
\aeq \bl{2\p_\nu \p_\ga h_{\mu\be} \p^\ga h^{\be\nu}}+2\p_\ga h_{\mu\be} \p^\ga (\p h)^{\be} \no\\
&&+2\p_\nu h_{\mu\be} \Box h^{\be\nu}+2h_{\mu\be} \Box (\p h)^{\be}
+2(\p h)_{\be} \Box h^{\be}_{\ \mu} \no\\
&&+2h^{\nu}_{\ \be}\p_\nu \Box h^{\be}_{\ \mu}, \\
(\p \bm{\chi}_{[8]})_\mu \aeq  -2\p_\nu \p_\ga h_\mu^{\ \dl}\p_\dl h^{\nu\ga}-2\p_\ga h_\mu^{\ \dl}\p_\dl (\p h)^{\ga}
+2\p_\nu \p_\ga h_{\mu\be}\p^\nu h^{\be\ga} \no\\
&&+2 \p_\ga h_{\mu\be}\Box h^{\be\ga}
+2\p_\ga (\p h)_{\be}\p_\mu h^{\be\ga}+2\p_\ga h^{\nu}_{\ \be}\p_\nu \p_\mu h^{\be\ga} \no\\
&&+2\p_\nu h_{\mu\be}\p^\nu (\p h)^{\be} +2 h_{\mu\be}\Box (\p h)^{\be}
+2 (\p h)_{\be}\p_\mu (\p h)^{\be} \no\\
&&+2 h^{\nu}_{\ \be}\p_\nu \p_\mu (\p h)^{\be} ,\\
(\p \bm{\chi}_{[9]})_\mu
 \aeq 2\p_\mu \p_\ga h_{\al\be} \p^\al h^{\be\ga}+2 \p_\ga h_{\al\be} \p_\mu\p^\al h^{\be\ga}
+2\p_\mu h^{\al\be} \p_\al (\p h)_\be \no\\
&& +2 h^{\al\be} \p_\mu \p_\al (\p h)_\be
-\p_\nu \p_\ga h \p_\mu h^{\nu\ga}- \p_\ga h \p_\mu (\p h)^{\ga} \no\\
&&-\p_\ga h \Box h_\mu^{\ \ga} +\p_\nu (\p h)_{\mu}\p^\nu h+(\p h)_{\mu}\Box h
+(\p \p h)\p_\mu h \no\\
&&+2(\p h)^{\nu}\p_\nu \p_\mu h
+h^{\al}_{\ \mu}\p_\al \Box h  
 +h^{\al\nu}\p_\nu \p_\al \p_\mu h ,\\
(\p \bm{\chi}_{[10]})_\mu \aeq 4 \p_\ga h_{\al\be}\p_\mu \p^\ga h^{\al\be}+2\p_\mu h_{\al\be}\Box h^{\al\be}
+2 h_{\al\be}\p_\mu \Box h^{\al\be} \no\\
&&+2(\p h)_\mu\Box h+2h^{\nu}_{\ \mu}\p_\nu \Box h ,\\
(\p \bm{\chi}_{[11]})_\mu 
\aeq 4\p^\nu h \p_\nu \p_\mu h+ 2 \p_\mu h\Box h +2h\p_\mu \Box h \no\\
&& + 2\p_\mu h (\p \p h) + 2h\p_\mu  (\p \p h) , \\
(\p \bm{\chi}_{[12]})_\mu 
\aeq -\p_\mu (\p h)^{\ga}(\p h)_{\ga}
-\Box h_\mu^{\ \ga}(\p h)_{\ga}-\p^\nu h_\mu^{\ \ga} \p_\nu (\p h)_{\ga} \no\\
&&+3\p_\nu (\p h)_\mu (\p h)^\nu 
+2 (\p h)_\mu (\p \p h) 
+\p_\nu h_\mu^{\ \be}\p_\be (\p h)^{\nu} \no\\
&&+ h_\mu^{\ \be}\p_\be (\p \p h)
+ h^{\nu\be}\p_\nu \p_\be (\p h)_{\mu} 
+2\p_\nu \p_\mu h^{\al\be}\p_\be h^\nu_{\ \al} \no\\
&&+\Box h^{\al\be}\p_\be h_{\mu\al} 
+2\p^\nu h^{\al\be}\p_\nu \p_\be h_{\mu\al}
+ h^{\al\be}\p_\be \p_\mu (\p h)_{\al} \no\\
&&+ h^{\al\be}\p_\be \Box h_{\mu\al} ,\\
(\p \bm{\chi}_{[13]})_\mu 
\aeq 4(\p h)^{\al}\p_\mu (\p h)_{\al}
+2\p_\mu h^{\al\be} \p_\al (\p h)_{\be}
+2 h^{\al\be} \p_\mu \p_\al (\p h)_{\be} \no\\
&&-\p_\nu \p_\mu h (\p h)^{\nu}
-\p_\mu h (\p \p h)
-\Box h (\p h)_{\mu} \no\\
&&-\p^\nu h \p_\nu (\p h)_{\mu}
+\p_\mu (\p h)^{\al}\p_\al h 
+\p_\mu h^{\al\nu}\p_\nu \p_\al h \no\\
&&+\Box h^{\al}_{\ \mu}\p_\al h 
+2\p^\nu h^{\al}_{\ \mu}\p_\nu \p_\al h 
+(\p h)^{\al}\p_\mu \p_\al h \no\\
&&+h^{\al\nu}\p_\nu \p_\mu \p_\al h
+ h^{\al}_{\ \mu}\Box \p_\al h,\\
(\p \bm{\chi}_{[14]})_\mu 
\aeq 2(\p h)_\mu(\p \p h) + 2h_\mu^{\ \nu}\p_\nu(\p \p h)
+4\p^\nu h^{\al\be} \p_\nu \p_\mu h_{\al\be} \no\\
&&+2\p_\mu h_{\al\be}\Box h^{\al\be}
+2h_{\al\be}\p_\mu \Box h^{\al\be} ,\\
(\p \bm{\chi}_{[15]})_\mu 
\aeq -4(\p h)_{\al}\p_\mu (\p h)^{\al}+2\p_\nu \p_\mu h(\p h)^{\nu}+2\p_\mu h(\p \p h) \no\\
&&+2\Box h(\p h)_{\mu}+4\p^\nu h\p_\nu (\p h)_{\mu}
+2\p_\nu h\p_\mu(\p h)^{\nu} \no\\
&&+2 h\p_\mu(\p \p h) +2 h\Box (\p h)_{\mu} ,\\
(\p \bm{\chi}_{[16]})_\mu 
 \aeq -2 \p_\nu(\p h)_{\mu}(\p h)^{\nu}-2 (\p h)_{\mu}(\p \p h)
 +2\p_\mu h^{\nu\al} \p_\nu (\p h)_\al \no\\
&& +2\Box h_\mu^{\ \al}(\p h)_\al 
 +4\p^\nu h_\mu^{\ \al}\p_\nu (\p h)_\al
+4 (\p h)^{\al}\p_\mu(\p h)_\al \no\\
&&+2 h^{\nu\al}\p_\nu \p_\mu(\p h)_\al
+2 h_\mu^{\ \al}\Box(\p h)_\al.
\eea
The solution of \re{cond_2_re} is given by \cite{GS}
\bea
\bL^{(3)} \aeq \bL_\RM{E}^{(3)}+x\Big([3]+[9]-[13]-[15]\Big)+y\Big([6]+[8]-[12]-[16]\Big)
\eea
where $x$ and $y$ are arbitrary real constants and 
\bea
\bL_\RM{E}^{(3)}/g \aeq \half[1]-\half[2]+[3]-[4]+[5]-4[6]+2[7]-2[8] +2[9] \no\\
&&-2[10]-[11]+ 2[13]+ 2[14] . \la{EH_3_0}
\eea
Because of \re{weak}, $\bL^{(3)} \we \bL_\RM{E}^{(3)}$ holds.
We derive \re{EH_3_0} in \res{A_B_E}.

\subsection{Feynman's cubic Lagrangian density} \la{s_4_4}

The expression given by Feynman \cite{Feynman} is 
\bea
\hs{-10mm}\bL^{(3)}=\bL^{(3)}_\RM{Feynman} \aeqd -g\Big[ h^{\al\be}\bar{h}^{\ga\dl}\p_\ga \p_\dl \bar{h}_{\al\be}+h_\ga^{\ \be}h^{\ga\al}\Box \bar{h}_{\al\be} 
-2h^{\al\be}h_{\be}^{\ \dl}\p_\ga \p_\dl \bar{h}^\ga_{\ \al}  \no\\
&&+2 \bar{h}_{\al\be}(\p \bar{h})^\al (\p \bar{h})^\be + \half h_{\al\be}h^{\al\be}\p_\ga \p_\dl \bar{h}^{\ga\dl} + \f{1}{4}hh\p_\ga \p_\dl \bar{h}^{\ga\dl} \Big], \la{L_3_F_0}
\eea
where 
\bea
\bar{h}_{\mu\nu} \aeqd h_{\mu\nu}-\half \eta_{\mu\nu}h \com (\p\bar{h})^\mu \defe \p_\nu\bar{h}^{\nu\mu} .
\eea
$\bL^{(3)}_\RM{Feynman}$ can be rewritten as (\res{A_B_F})
\bea
\bL^{(3)}_\RM{Feynman}/g 
\aeqw \half[1]-\half[2]+[3]-[4]+[5]-4[6]+2[7]-4[8]+2[9] \no\\
&&-2[10] -[11]+2[12]+2[13]+2[14] \la{L_3_F_2}.
\eea
Because of
\bea
\bL^{(3)}_\RM{Feynman}/g - \bL_\RM{E}^{(3)}/g \aeqw -2[8]+2[12] \stackrel{\mathrm{w}}{\ne} 0, \la{deff_F_E}
\eea
$\bL^{(3)}_\RM{Feynman}$ is not a solution of \re{cond_2_re} in the present framework.

\rd{In our framework, \re{cond_2_re} must be satisfied at the Lagrangian density level.
$\bL^{(3)}_\RM{Feynman}$ does not satisfy it, however under certain backgrounds, the contribution vanishes and the observables may agree. 
For instance, in a spherically symmetric and static system, the perihelion shift agrees (\res{s_6}). 
The difference is expected to matter, e.g. in genuinely time-dependent situations (radiation) or in processes sensitive to the full off-shell cubic vertex such as scattering amplitudes, as well as in higher post-Newtonian orders beyond the restricted static sector tested by perihelion precession.}

\rd{
This note adopts the identification $g_{\mu\nu}=\eta_{\mu\nu}+h_{\mu\nu}$ and the equation \re{cond_2} based on it.
A broader equivalence encompassing local field redefinition requires separate discussion.
}

\section{Fourth-order Lagrangian density} \la{A_4}

\rd{The candidate of $\bL^{(4)}$ is the sum of terms that are quadratic in $h_{\al\be}$ and quadratic in $\p_\ga h_{\al\be}$. 
}
$\bL^{(4)}$ is given by $\bL^{(4)}=\sum_{n=1}^{43} \bl{c_n}\{n\}$ with
\bea
&&\hs{-10mm}\{1\}= h^{\al\be} h^{\ga\dl} \p_{\be}h_{\dl\ep} \p_{\ga}h_{\al}{}^{\ep},\
\{2\}= h_{\al}{}^{\ga} h^{\al\be} \p_{\be}h^{\dl\ep} \p_{\ga}h_{\dl\ep}, \
\{3\}=h h^{\be\ga} \p_{\be}h^{\dl\ep} \p_{\ga}h_{\dl\ep} ,\no\\
&&\hs{-10mm}\{4\}=h_{\al}{}^{\ga} h^{\al\be} \p_{\be}h \p_{\ga}h , \
\{5\}=h h^{\al\be} \p_{\al}h \p_{\be}h , \
\{6\}=h_{\al}{}^{\ga} h^{\al\be} \p_{\ga}h (\p h)_{\be} ,\no\\
&&\hs{-10mm}\{7\}= hh^{\al\be} \p_{\al}h  (\p h)_{\be} , \
\{8\}=h^{\al\be} h^{\ga\dl} \p_{\ga}h_{\al}{}^{\ep} \p_{\dl}h_{\be\ep} , \
\{9\}=h^{\al\be} h^{\ga\dl} \p_{\be}h_{\al}{}^{\ep} \p_{\dl}h_{\ga\ep},\no\\
&&\hs{-10mm}\{10\}= h^{\al\be} h^{\ga\dl} \p_{\be}h_{\al\ga} \p_{\dl}h , \
\{11\}= h^{\al\be} h^{\ga\dl} \p_{\ga}h_{\al\be} \p_{\dl}h , \
\{12\}=h_{\al}{}^{\ga} h^{\al\be} \p_{\ga}h_{\be}{}^{\dl} \p_{\dl}h ,\no\\
&&\hs{-10mm}\{13\}= h h^{\be\ga} \p_{\ga}h_{\be}{}^{\dl} \p_{\dl}h , \
\{14\}=h_{\al}{}^{\ga} h^{\al\be} \p_{\dl}h \p^{\dl}h_{\be\ga} , \
\{15\}= h h^{\be\ga} \p_{\dl}h \p^{\dl}h_{\be\ga} ,\no\\
&&\hs{-10mm}\{16\}=(h^2) \p_{\al}h \p^{\al}h , \ 
\{17\}=h^2 \p_{\al}h \p^{\al}h , \
\{18\}=h_{\al}{}^{\ga} h^{\al\be} (\p h)_\be (\p h)_\ga ,\no\\
&&\hs{-10mm}\{19\}=h h^{\al\be} (\p h)_\al (\p h)_\be , \
\{20\} =h^{\al\be} h^{\ga\dl} \p_{\be}h_{\al\ga} (\p h)_\dl , \
\{21\} = h^{\al\be} h^{\ga\dl} \p_{\ga}h_{\al\be} (\p h)_\dl ,\no\\
&&\hs{-10mm}\{22\}=h_{\al}{}^{\ga} h^{\al\be} \p_{\ga}h_{\be}{}^{\dl} (\p h)_\dl , \
\{23\}=h h^{\be\ga} \p_{\ga}h_{\be}{}^{\dl} (\p h)_\dl , \
\{24\}= (h^2)(\p h)_\al(\p h)^\al ,\no\\
&&\hs{-10mm}\{25\} = h^2 (\p h)_\al(\p h)^\al , \
\{26\}=h_{\al}{}^{\ga} h^{\al\be} \p^{\dl}h_{\be\ga} (\p h)_\dl , \
\{27\}= hh^{\be\ga} \p^{\dl}h_{\be\ga} (\p h)_\dl ,\no\\
&&\hs{-10mm}\{28\}= (h^2) \p^{\al}h (\p h)_\al , \
\{29\}= h^2\p^{\al}h (\p h)_\al , \
\{30\}=h^{\al\be} h^{\ga\dl} \p_{\dl}h_{\ga\ep} \p^{\ep}h_{\al\be} ,\no\\
&&\hs{-10mm}\{31\}= h^{\al\be} h^{\ga\dl} \p_{\ep}h_{\ga\dl} \p^{\ep}h_{\al\be} , \
\{32\}= h^{\al\be} h^{\ga\dl} \p_{\dl}h_{\be\ep} \p^{\ep}h_{\al\ga} , \
\{33\}= h^{\al\be} h^{\ga\dl} \p_{\ep}h_{\be\dl} \p^{\ep}h_{\al\ga} ,\no\\
&&\hs{-10mm}\{34\}= h_{\al}{}^{\ga} h^{\al\be} \p_{\ga}h_{\dl\ep} \p^{\ep}h_{\be}{}^{\dl} , \
\{35\}= h h^{\be\ga} \p_{\ga}h_{\dl\ep} \p^{\ep}h_{\be}{}^{\dl} , \
\{36\}=h_{\al}{}^{\ga} h^{\al\be} \p_{\dl}h_{\ga\ep} \p^{\ep}h_{\be}{}^{\dl} ,\no\\
&&\hs{-10mm}\{37\}= hh^{\be\ga} \p_{\dl}h_{\ga\ep} \p^{\ep}h_{\be}{}^{\dl} , \
\{38\}= h_{\al}{}^{\ga} h^{\al\be} \p_{\ep}h_{\ga\dl} \p^{\ep}h_{\be}{}^{\dl} , \
\{39\}=h h^{\be\ga} \p_{\ep}h_{\ga\dl} \p^{\ep}h_{\be}{}^{\dl} ,\no\\
&&\hs{-10mm}\{40\}= (h^2) \p_{\dl}h_{\ga\ep} \p^{\ep}h^{\ga\dl}  , \
\{41\}= h^2 \p_{\dl}h_{\ga\ep} \p^{\ep}h^{\ga\dl} , \
\{42\}=(h^2) \p_{\ep}h_{\ga\dl} \p^{\ep}h^{\ga\dl},\no\\
&&\hs{-10mm}\{43\}=h^2 \p_{\ep}h_{\ga\dl} \p^{\ep}h^{\ga\dl} ,
\eea
where $(h^2)\defe h_{\mu\nu}h^{\mu\nu}$. 
\rd{The following relations hold:
\bea
\{1\}-\{9\}-\{20\}+\{22\}+\{32\}-\{34\} \aeqw 0 ,\la{we_0_1}\\
-\{6\} + \{12\}-\{19\}-\{23\}+\{35\}+\{37\} \aeqw 0 ,\\
-\{7\}+\{13\}-\half \{25\}+\half \{41\} \aeqw 0 ,\\
\{18\}+\{20\}+\{22\}-\{32\}-\{34\}-\{36\} \aeqw 0,\\
\{21\}+\half \{24\}-\{30\}-\half \{40\} \aeqw 0. \la{we_0_5}
\eea
}
The solution of \re{cond_3} is given by (we used the Wolfram Language with the xAct package)
\bea
&&\hs{-10mm}4\ka \bm{\mL}^{(4)} \no\\
\aeq  C_{1}^{} \{1\}
 - \f{1}{2} \{2\}
  + \f{1}{4} \{3\}
  + \f{1}{2} \{4\}
  - \f{1}{4} \{5\}
+ (-2 - C_{12}^{}) \{6\}
+ (1 - C_{13}^{})\{7\} \no\\
&&- \{8\}
+ (1 - C_{1}^{}) \{9\} 
 - \{10\}
  + \{11\}
  + C_{12}^{} \{12\} 
  + C_{13}^{} \{13\}
  + \{14\} \no\\
 && - \half \{15\}
  - \f{1}{8}\{16\}
  +\f{1}{16}\{17\}
  + C_{18}^{} \{18\} 
   + (-1 - C_{12}^{})\{19\} \no\\
&&  + (1 - C_{1}^{} + C_{18}^{})\{20\}
 + C_{21}^{} \{21\} 
+ (-1 + C_{1}^{} + C_{18}^{}) \{22\} \no\\
&&+ (-1 - C_{12}^{}) \{23\}
+ \Big(\f{1}{2} + \f{1}{2} C_{21}^{}\Big) \{24\} 
+ \Big(\f{1}{4} - \f{1}{2} C_{13}^{}\Big) \{25\} 
 -\{26\} \no\\
&&+ \half \{27\} 
+ \f{1}{4} \{28\} 
- \f{1}{8} \{29\} 
+ (-2 - C_{21}^{})\{30\} 
+ \f{1}{2} \{31\} \no\\
&&+ (1 + C_{1}^{} - C_{18}^{}) \{32\} 
- \f{1}{2} \{33\} 
+ (3 - C_{1}^{} - C_{18}^{}) \{34\} 
+ C_{12}^{} \{35\} \no\\
&&+ (1 - C_{18}^{}) \{36\} 
+ \Big(\f{1}{2} + C_{12}^{}\Big) \{37\} 
 - \{38\} 
 + \f{1}{2} \{39\} \no\\
&& +\Big(- \f{3}{4} - \f{1}{2} C_{21}^{}\Big)\{40\}  
 + \Big(- \f{1}{8} + \f{1}{2} C_{13}^{}\Big) \{41\} 
 + \f{1}{8} \{42\} 
  - \f{1}{16} \{43\}  . \la{general_43}
\eea
Here, $C_1$, $C_{12}$, $C_{13}$, $C_{18}$, and $C_{21}$ are arbitrary real constants. 
$\bm{\chi}_{(3)}^{\mu\nu}$ does not depend on these constants. 
$\bm{\mL}_\RM{E}^{(4)}$ is given by \cite{L4}
\bea
&&\hs{-20mm}4\ka \bm{\mL}_\RM{E}^{(4)} 
= -\Big(h^2-2 (h^2) \Big)
\Big(\frac{1}{16} \p^\sig h^{\gamma \delta} \p_\sig h_{\gamma \delta}
-\frac{1}{8}\p^\sig h^{\gamma \delta } \p_\dl h_{\gamma \sigma}
+\frac{1}{8} \p_\dl h (\p h)^{\delta}
-\frac{1}{16} \p_\dl h \p^\dl h\Big) \no\\
&&-h h^{\beta \gamma}
\Big(-\frac{1}{2}\p_\dl h_{\beta \gamma}(\p h)^{\delta}
+\frac{1}{2} \p_\dl h_{\beta \gamma} \p^\dl h
+\frac{1}{4} \p_\be h \p_\ga h
-\half  \p_\be h (\p h)_\ga  \no\\
&&+\p_\sig h_{\beta }^\delta \p_\ga h_{\dl}^\sigma
-\frac{1}{4} \p_\be h^{\delta \sigma}\p_\ga h_{\delta \sigma }
-\frac{1}{2} \p_\sig h_{\beta}^\delta \p^\sig h_{\delta \gamma}
-\frac{1}{2} \p_\dl h \p_\ga h_{\beta}^\delta 
+\frac{1}{2} \p_\sig h_{\beta \delta} \p^\dl h_\gamma^{\sigma} \Big) \no\\
&&-h_{\ \beta}^\alpha h^{\beta \gamma}
\Big(\p_\sig h \p_\ga h_{\alpha }^\sigma
-\p_\dl h_{\alpha \gamma} \p^\dl h
+\frac{1}{2} \p_\al h^{\delta \sigma} \p_\ga h_{\delta \sigma}
 -\p_\sig h_{\alpha}^\delta \p_\dl h_{\gamma}^\sigma \no\\
&& -2 \p_\sig h_{\alpha}^\delta \p_\ga h_{\delta}^\sigma
+\p_\dl h_{\alpha \gamma} (\p h)^\dl
+\p_\al h (\p h)_{\gamma}
-\frac{1}{2} \p_\al h \p_\ga h 
+\p_\sig h_{\alpha}^\delta \p^\sig h_{\gamma \delta} \Big) \no\\
&&-h^{\alpha \gamma} h^{\beta \delta}
\Big(\p_\be h_{\alpha \gamma} (\p h)_\dl 
-\p_\dl h_{\alpha \gamma} \p_\be h
+\frac{1}{2} \p_\sig h_{\alpha \beta} \p^\sig h_{\gamma \delta}
-\frac{1}{2} \p_\sig h_{\alpha \gamma} \p^\sig h_{\beta \delta}\no\\
&&+\p_\be h_{\alpha}^\sigma \p_\dl h_{\gamma \sigma}
-\p_\be h_{\alpha}^\sigma \p_\ga h_{\delta \sigma}
+\p_\dl h_{\alpha \beta} \p_\ga h
-2\p_\be h_{\alpha}^\sigma \p_\sig h_{\delta \gamma}
+\p_\sig h_{\alpha \gamma} \p_\dl h_{\beta}^\sig \Big) \no\\
\aeq \{1\}-\half \{2\}+\f{1}{4}\{3\}+\half \{4\}-\f{1}{4}\{5\}
-\{6\}+\ \half \{7\}-\{8\}- \{10\} \no\\
&&+\{11\}-\{12\}+\half \{13\}+\{14\}-\half \{15\}
-\f{1}{8}\{16\}+\f{1}{16}\{17\}\no\\
&&-\{21\}-\{26\}+\half\{27\}+\f{1}{4}\{28\}-\f{1}{8}\{29\}-\{30\} \no\\
&&+\half\{31\}+2\{32\}-\half\{33\}+2\{34\}-\{35\}+\{36\} \no\\
&&-\half\{37\}-\{38\}+\half\{39\} -\f{1}{4}\{40\} 
+\f{1}{8}\{41\}+\f{1}{8}\{42\}-\f{1}{16}\{43\} .
\eea
Reference \cite{L4} contains a single error in the term $\{7\}/2$. 
The above expression is obtained by substituting 
\bea
C_1 =1 \com
C_{12} = -1 \com
C_{13} = \half \com  
C_{18} = 0 \com
C_{21} = -1 
\eea
into \re{general_43}. 
\rd{
From this and \re{we_0_1}-\re{we_0_5}, $\bm{\mL}^{(4)} \we \bm{\mL}_\RM{E}^{(4)} $ holds. 
}

\section{Perihelion shift} \la{s_6}

In this section, we consider a spherically symmetric and static system. 
We examine the motion of a particle around a star and investigate the perihelion shift.
The second-order Lagrangian density $\bL^{(2)}$ alone cannot account for the observed perihelion shift; 
to correctly determine the perihelion shift, it is necessary to consider the third-order Lagrangian density $\bL^{(3)}$.

The equation of motion \re{p_eom} can be rewritten as
\bea
\f{d}{d\tau}\Big[(\eta_{\sig\nu}+h_{\sig\nu})\f{dx^\sig}{d\tau}\Big]=\half \p_\nu h_{\mu\sig}\f{dx^\mu}{d\tau}\f{dx^\sig}{d\tau} .\la{p_eom_2}
\eea
We suppose that
\bea
h_{\mu\nu} \aeq \RM{diag}(h_0,h_s,h_s,h_s).
\eea
Then, the spatial components ($i=1,2,3$) of \re{p_eom_2} become
\bea
\f{d}{d\tau}\Big[(1+h_s)\dot{x}^i \Big] \aeq \half \Big[\p_i h_0 \dot{t}^2+\p_i h_s (\dot{x}^2+\dot{y}^2+\dot{z}^2) \Big], \la{p_eom_s}
\eea
where $\dot{X}\defe dX/d\tau$ and $t=x^0$. 
The time component of \re{p_eom_2} become
\bea
\f{d}{d\tau}\Big[(1-h_0)\dot{t}\Big] \aeq 0. \la{p_eom_t}
\eea
We used $\p_0 h_{\mu\nu} =0$. 
In this case, \re{e_a_eom} becomes
\bea
(1-h_0)\dot{t}^2-(1+h_s)(\dot{x}^2+\dot{y}^2+\dot{z}^2) \aeq 1. \la{tau_cond}
\eea
From \re{p_eom_t}, we have
\bea
K\defe (1-h_0)\dot{t} = \const.
\eea
Using this equation and \re{tau_cond}, we have
\bea
\f{K^2}{1-h_0}-(1+h_s)(\dot{x}^2+\dot{y}^2+\dot{z}^2) \aeq 1. \la{eq_6.7}
\eea
$h_0$ and $h_s$ depend on only $r\defe \sqrt{x^2+y^2+z^2}$. 
Thus, using \re{p_eom_s}, we have
\bea
\f{d}{d\tau}\Big[(1+h_s)(\dot{x}^i x^k-\dot{x}^k x^i) \Big] 
\aeq \f{d}{d\tau}\Big[(1+h_s)\dot{x}^i \Big] x^k-\f{d}{d\tau}\Big[(1+h_s)\dot{x}^k \Big]x^i \no\\
\aeq 0.
\eea
Using this equation, 
\bea
L_1 \aeqd (1+h_s)(\dot{z} y-\dot{y}z) \com
L_2 \defe (1+h_s)(\dot{x} z-\dot{z}x) , \no\\
L \aeqd (1+h_s)(\dot{y} x-\dot{x}y) 
\eea
are conserved. 
Setting $L_1 = L_2 = 0$ confines the motion to the equatorial plane, $\varphi = \pi/2$ 
($x=r\sin \varphi \cos \theta$, $y=r\sin \varphi \sin \theta$, and $z=r\cos \varphi$). 
Then, we have 
\bea
L \aeq (1+h_s)r^2\dot{\theta} \la{L=} ,
\eea
and $\dot{x}^2+\dot{y}^2+\dot{z}^2 =r^2\dot{\theta}^2+\big(\f{dr}{d\theta}\big)^2\dot{\theta}^2 $.
\re{eq_6.7} becomes
\bea
\f{K^2}{1-h_0}-(1+h_s)\dot{\theta}^2 \Big[r^2+\Big(\f{dr}{d\theta}\Big)^2 \Big] \aeq 1.
\eea
Using $\dot{\theta} = \f{L}{(1+h_s)r^2}$ because of \re{L=}, the above equation becomes
\bea
\f{K^2}{1-h_0}-\f{L^2}{(1+h_s)r^4}\Big[r^2+\Big(\f{dr}{d\theta}\Big)^2 \Big] \aeq 1.
\eea
We define $u \defe1/r$.
Then, the above equation becomes
\bea
u^2+\Big(\f{du}{d\theta}\Big)^2 \aeq \Big(\f{K^2}{1-h_0}-1 \Big)\f{1+h_s}{L^2}. \la{eq_u_pre}
\eea
Here, we assume that
\bea
h_0 \aeq -\al \phi-a \phi^2 +O(\phi^3) ,\\
h_s \aeq -\be \phi-b \phi^2 +O(\phi^3), 
\eea
where $\phi \defe -2G_\RM{N}Mu$. 
Here, $M$ is the mass of the star and $G_\RM{N}$ is the universal gravitational constant.
Then, we have
\bea
\Big(\f{K^2}{1-h_0}-1 \Big)\f{1+h_s}{L^2} \aeq A+Bu+Cu^2+O(u^3), \la{ABC}
\eea
where
\bea
A \aeq \f{K^2-1}{L^2} \com
B = \f{2G_\RM{N}M}{L^2}\Big[K^2\al+(K^2-1)\be \Big] ,\no\\
C \aeq \f{(2G_\RM{N}M)^2}{L^2} \Big[ K^2(\al^2+\al\be-a)-(K^2-1)b \Big].
\eea
Substituting \re{ABC} into \re{eq_u_pre}, we have
\bea
u^2+\Big(\f{du}{d\theta}\Big)^2 \aeq A+Bu+Cu^2.
\eea
Here, we ignored the term $O(u^3)$. 
Differentiating the above equation with respect to $\theta$, we have 
\bea
\f{d^2u}{d\theta^2} \aeq \half B -(1-C)u.
\eea
Putting $u=:\f{B}{2(1-C)} +v$, the above equation becomes
\bea
\f{d^2v}{d\theta^2} \aeq -(1-C)v .
\eea
The solution is given by 
\bea
v \aeq v_0\cos (\sqrt{1-C} \theta)+v_1 \sin (\sqrt{1-C} \theta).
\eea
Thus, the precession of the perihelion point over one cycle $\dl$ is given by
\bea
\dl \aeq \f{2\pi}{\sqrt{1-C}}-2\pi =C\pi +O(C^2) \no\\
\aeqap \pi\f{(2G_\RM{N}M)^2}{L^2} \Big[ K^2(\al^2+\al\be-a)-(K^2-1)b \Big] \no\\
\aeqap \pi\f{(2G_\RM{N}M)^2}{L^2} (\al^2+\al\be-a) \la{saisa}.
\eea
We used $K^2 \approx 1$. 

We consider the Lagrangian density of the gravitational field up to third order. 
The total action is given by 
\bea
S_\tot \aeq S^{(2)}+S^{(3)}+\tl S_\RM{particle} + \int d^4x \ \half h_{\mu\nu}(x)\bm{T}^{\mu\nu}_\RM{(p)}(x).
\eea
In this case, the Euler-Lagrange equation of the gravitational field is given by
\bea
\bm{\chi}_{(1)}^{\mu\nu}[h]+\bm{\chi}_{(2)}^{\mu\nu}[h] \aeq \bm{T}^{\mu\nu}_\RM{(p)}.
\eea
We expand $h_{\mu\nu}$ as $h_{\mu\nu} = h_{\mu\nu}^{\langle 1 \rangle}+h_{\mu\nu}^{\langle 2 \rangle} +\cdots$ 
where $h_{\mu\nu}^{\langle n \rangle}$ is $n$-th-order term in $G_\RM{N}$.
We have
\bea
\bm{\chi}_{(1)}^{\mu\nu}[h^{\langle 1 \rangle}] \aeq \bm{T}^{\mu\nu}_\RM{(p)} ,\la{eq_grav_1}\\
\bm{\chi}_{(1)}^{\mu\nu}[h^{\langle 2 \rangle}] \aeq -\bm{\chi}_{(2)}^{\mu\nu}[h^{\langle 1 \rangle}] .\la{eq_grav_2}
\eea
By solving \re{eq_grav_1}, we have
\bea
(\al_{\langle 1 \rangle} , \be_{\langle 1 \rangle}, a_{\langle 1 \rangle}, b_{\langle 1 \rangle} )=(1,1,0,0).\la{al_be_a_b_1}
\eea
$ h_{\mu\nu}^{\langle 1 \rangle}$ is the solution obtained 
when considering the Lagrangian density of the gravitational field up to the second-order. 
Solving \re{eq_grav_2} yields $h_{\mu\nu}^{\langle 1 \rangle}+h_{\mu\nu}^{\langle 2 \rangle}$, which gives 
\bea
(\al_{\langle 2 \rangle} , \be_{\langle 2 \rangle}, a_{\langle 2 \rangle}, b_{\langle 2 \rangle})=\Big(1,1,\half, -\f{3}{8}\Big).
\eea
\rd{This value agrees with Ortín \cite{GS} and Nikishov \cite{Nikishov}. For comparison, see also the discussion in Feynman's lectures \cite{Feynman}.} 
Thus, we have
\bea
\dl_{\langle 1\rangle} \aeq \pi\f{(2G_\RM{N}M)^2}{L^2} \cdot 2=\f{4}{3}\dl_{\langle 2\rangle} ,\\
\dl_{\langle 2\rangle} \aeq \pi\f{(2G_\RM{N}M)^2}{L^2} \cdot \f{3}{2} .
\eea
$\dl_{\langle 2\rangle}$ agrees with the experiment, but $\dl_{\langle 1\rangle}$ does not. 
Because $\bm{\chi}_{[8]}^{\mu\nu}[h^{\langle 1 \rangle}]=\bm{\chi}_{[12]}^{\mu\nu}[h^{\langle 1 \rangle}]$ holds \cite{Nikishov} 
in this case, $\bL^{(3)}_\RM{Feynman}$ also gives the correct perihelion shift.

\section*{Note added in arXiv v3}

According to Ref.\cite{H}, the cubic Lagrangian density given by Feynman is 
\bea
\bL^{(3)} \aeq -g\Big[ h^{\al\be}\bar{h}^{\ga\dl}\p_\ga \p_\dl \bar{h}_{\al\be}+h_\ga^{\ \be}h^{\ga\al}\Box \bar{h}_{\al\be} 
-2h^{\al\be}h^{\ga\dl}\p_\ga \p_\be \bar{h}_{\dl\al}  \no\\
&&+2 \bar{h}_{\al\be}(\p \bar{h})^\al (\p \bar{h})^\be + \half h_{\al\be}h^{\al\be}\p_\ga \p_\dl \bar{h}^{\ga\dl} + \f{1}{4}hh\p_\ga \p_\dl \bar{h}^{\ga\dl} \Big]. 
\eea
The third term differs from \re{L_3_F_0}. 
This Lagrangian density is correct ($\bL^{(3)} \we \bL_\RM{E}^{(3)}$).

\appendix

\section{Third-order Lagrangian densities} \la{A_B}

\subsection{Expansion of Einstein Lagrangian density} \la{A_B_E}

We calculate $\bL^{(3)}_\RM{E}$. 
Putting $S \defe \sqrt{-\det(g_{\mu\nu})}$, we have $\bm{\mL}_\RM{E} =\f{1}{2\ka}SG $.
We expand $g^{\mu\nu}$ and $S$ as
\bea
g^{\mu\nu} \aeq \eta^{\mu\nu}+g^{\mu\nu}_{(1)}+g^{\mu\nu}_{(2)}+\cdots ,\\
S \aeq 1+ S^{(1)} +S^{(2)}+\cdots,
\eea
where $(n)$ represents the $n$-th-order term in $h_{\mu\nu}$. 
Using 
\bea
(A+B)^{-1} \aeq A^{-1}-A^{-1}BA^{-1}+A^{-1}BA^{-1}BA^{-1}-\cdots
\eea
for square matrices $A$ and $B$, we have
\bea
g^{\mu\nu}_{(1)} \aeq -h^{\mu\nu} \com g^{\mu\nu}_{(2)} = h^\mu_{\ \rho}h^{\rho\nu} .
\eea
Using $\det(A) =\exp \tr \ln A $, we have
\bea
\det(A+B) \aeq \det(A)\det(1+A^{-1}B) \no\\
\aeq \det(A)\exp \tr \ln (1+A^{-1}B) \no\\
\aeq \det(A)\Big(1+ \tr [A^{-1}B]-\half \tr[A^{-1}BA^{-1}B] \no\\
&&+\half \big(\tr [A^{-1}B] \big)^2 +\cdots \Big). \la{det_A+B}
\eea
The above equation leads to
\bea
\sqrt{-\det(A+B)} \aeq \sqrt{-\det(A)}\Big(1+ \half \tr [A^{-1}B] \no\\
&&+\f{1}{8}\big\{ \big(\tr [A^{-1}B] \big)^2-2\tr[A^{-1}BA^{-1}B] \big\}+\cdots \Big).
\eea
Thus, we have
\bea
S^{(1)} \aeq \half h^\mu_{\ \mu} = \half h \com
S^{(2)}= \f{1}{8}\Big( h^2 - 2 h^\mu_{\ \nu}h^{\nu}_{\ \mu}\Big).
\eea
$\bL^{(3)}_\RM{E}$ is given by
\bea
2\ka \bm{\mL}_\RM{E}^{(3)} \aeq G^{(3)}+S^{(1)}G^{(2)} = G^{(3)}+\half h G^{(2)} \la{L_3_A1}
\eea
where 
\bea
G^{(3)} \aeq G^{(3a)}+G^{(3b)} ,\\
G^{(3a)} \aeqd \eta^{\mu\nu}\Big[{}^{(2)}\Ga^\rho_{\ \ga\nu}{}^{(1)}\Ga^\ga_{\ \mu\rho}
+{}^{(1)}\Ga^\rho_{\ \ga\nu}{}^{(2)}\Ga^\ga_{\ \mu\rho} \no\\
&&-{}^{(2)}\Ga^\rho_{\ \ga\rho}{}^{(1)}\Ga^\ga_{\ \mu\nu}-{}^{(1)}\Ga^\rho_{\ \ga\rho}{}^{(2)}\Ga^\ga_{\ \mu\nu} \Big] \no\\
\aeqe \f{1}{4} \Big(\bL_1+\bL_2+\bL_3+\bL_4 \Big), \\
G^{(3b)} \aeqd -h^{\mu\nu}\Big[{}^{(1)}\Ga^\rho_{\ \ga\nu}{}^{(1)}\Ga^\ga_{\ \mu\rho}-{}^{(1)}\Ga^\rho_{\ \ga\rho}{}^{(1)}\Ga^\ga_{\ \mu\nu} \Big] 
=:  \f{1}{4} \Big(\bL_5+  \bL_6 \Big).
\eea
Here, ${}^{(n+1)}\Ga^\lm_{\ \mu\nu}=g^{\lm\rho}_{(n)}\Ga_{\rho\mu\nu}$ with $g^{\lm\rho}_{(0)}=\eta^{\lm\rho}$. 
Thus, we have
\bea
\bL_\RM{E}^{(3)}/g \aeq  4 G^{(3)}+2 h G^{(2)} =\sum_{k=1}^7 \bL_k
\eea
with $\bL_7  \defe 2 h G^{(2)}$ and $g = 1/(8\ka)$. 
$\{\bL_k\}_{k=1}^7$ are given by
\bea
\bL_1 \aeq [5]-2[6] \com
\bL_2 =[5]-2[6] \com
\bL_3 =  2[14]-[10] ,\no\\
\bL_4 \aeq 2[13]-[4] \com
\bL_5 =-[5]+2[7]-2[8] \com
\bL_6 = 2[9]-[10] ,\no\\
\bL_7 \aeq \half[1]-\half[2]+[3]-[11].
\eea
Then, we have
\bea
\bL_\RM{E}^{(3)}/g \aeq \half[1]-\half[2]+[3]-[4]+[5]-4[6]+2[7]-2[8] +2[9] \no\\
&&-2[10]-[11]+ 2[13]+ 2[14] .\la{EH_3}
\eea

\subsection{Derivation of \re{L_3_F_2}} \la{A_B_F}

Each term on the right-hand side of \re{L_3_F_0} is given by 
\bea
h^{\al\be}\bar{h}^{\ga\dl}\p_\ga \p_\dl \bar{h}_{\al\be} 
\aeqw - \p_\dl (h^{\al\be}\bar{h}^{\ga\dl}) \p_\ga \bar{h}_{\al\be} \no\\
\aeq -[5]+\half [2]+\half [4]-\half[1]-[14]\no\\
&&+\half[10]+\half[11] ,\\
h_\ga^{\ \be}h^{\ga\al}\Box \bar{h}_{\al\be} 
\aeqw -\p_\dl (h_\ga^{\ \be}h^{\ga\al})\p^\dl \bar{h}_{\al\be} \no\\
\aeq -2[7]+[10] , \\
-2h^{\al\be}h_{\be}^{\ \dl}\p_\ga \p_\dl \bar{h}^\ga_{\ \al} 
\aeqw 2\p_\ga (h^{\al\be}h_{\be}^{\ \dl}) \p_\dl \bar{h}^\ga_{\ \al} \no\\
\aeq 2[6]-[13]+2[8]-[9] , \\
2 \bar{h}_{\al\be}(\p \bar{h})^\al (\p \bar{h})^\be 
\aeq 2[16]-[15]-2[13]+[11]+\half [4]-\f{1}{4}[1] ,\\
 \half h_{\al\be}h^{\al\be}\p_\ga \p_\dl \bar{h}^{\ga\dl} 
 \aeqw - \half \p_\ga (h_{\al\be}h^{\al\be})\p_\dl \bar{h}^{\ga\dl} \no\\
 \aeq -[14]+\half [10],\\
\f{1}{4}hh\p_\ga \p_\dl \bar{h}^{\ga\dl} \aeqw -\half h\p_\ga h (\p h)^{\ga}+\f{1}{4} h\p_\ga h \p^\ga h \no\\
\aeq -\half [11]+\f{1}{4}[1].
\eea
Thus, we have
\bea
\bL^{(3)}_\RM{Feynman}/g \aeqw \half[1]-\half[2]-[4]+[5]-2[6]+2[7]-2[8]+[9] \no\\
&&-2[10]-[11]+3[13]+2[14]+[15]-2[16] .\la{L_3_F_1}
\eea
The above equation and \re{weak} lead to \re{L_3_F_2}.

\subsection{Other literature}

Reference \cite{Janssen} studied $\bL^{(3)}$ and obtained $\bL^{(3)}= \bL^{(3)}_\RM{E}$. 
Reference \cite{LP} calculated $\bL^{(3)}$ as in \res{A_3} and obtained 
\bea
\hs{-5mm}\bL^{(3)}=\bL^{(3)}_\RM{Lopez-Pinto}\aeqd g\Big(\half[1] - \half[2] - [4] + [5] - 4[6] + 2[7] - 2[8] + [9]  \no\\
&&- 2[10] -[11] + 3[13] + 2[14] + [15] \Big)\we \bL_\RM{E}^{(3)}.  \la{L_3_LP_1}
\eea

\end{document}